\documentclass{article}%
\usepackage[numbered,framed]{matlab-prettifier}
\usepackage{enumitem}
\usepackage{multicol}
\usepackage{fancyvrb}
\usepackage{pdfpages}
\usepackage{amsthm}
\usepackage{amsfonts}
\usepackage{amsmath}
\usepackage{graphicx}
\usepackage{amssymb}%
\newtheorem{theorem}{Theorem}

\newtheorem{conjecture}[theorem]{Conjecture}

\newtheorem{corollary}[theorem]{Corollary}
\newtheorem{criterion}[theorem]{Criterion}

\newtheorem{proposition}[theorem]{Proposition}
\newtheorem{remark}[theorem]{Remark}

\def\0{\mbox{\bf 0}}

\definecolor{darkgreen}{rgb}{0.1, 0.5, 0.1}
\begin{document}

\title{Approximation and decomposition of attractors of a Hopfield neural network system}
\author{Marius-F. Danca$^{1,2}$,\and\ Guanrong Chen$^{3}$
\and $^{1}$STAR-UBB Institute, Babes-Bolyai University, Cluj-Napoca, Romania,
\and $^{2}$Romanian Institute of Science and Technology, Cluj-Napoca, Romania,
\\
email: m.f.danca@gmail.com
\and $^{3}$Department of Electrical Engineering, City University of Hong Kong, P.R. China,
\\email: eegchen@cityu.edu.hk}
\maketitle

\begin{abstract}
 In this paper, the Parameter Switching (PS) algorithm is used to numerically approximate attractors of a Hopfield Neural Network (HNN) system. The PS algorithm is a convergent scheme designed for approximating the attractors of an autonomous nonlinear system, depending linearly on a real parameter. Aided by the PS algorithm, it is shown that every attractor of the HNN system can be expressed as a convex combination of other attractors. The HNN system can easily be written in the form of a linear parameter dependence system, to which the PS algorithm can be applied. This work suggests the possibility to use the PS algorithm as a control-like or anticontrol-like method for chaos.
\end{abstract}

\textbf{keywords}: Hopfield neural network system; Parameter switching algorithm; Numerical attractor; Attractors approximation, Attractor decomposition
\section{Introduction}

Most autonomous chaotic systems of integer order such as the Lorenz system, Roessler system, Chen system, etc., can be modeled by the following Initial Value Problem (IVP):
\begin{equation}\label{unu}
\dot{{x}}(t)=f_{p}({x(t)}),\quad{x}(0)={x}_{0},\quad t\in I:=[0,\infty),
\end{equation}
\noindent with $~{x}_{0}\in\mathbb{R}^{n}$, $f_{p}:\mathbb{R}^{n}\rightarrow\mathbb{R}^{n}$ a continuous function depending linearly on the bifurcation parameter $p \in\mathbb{R}$,
\begin{equation}\label{doi}
f_{p}({x)=}pB{x}+g({x)}.
\end{equation}
where $g:\mathbb{R}^{n}\rightarrow\mathbb{R}^{n}~\ $is a continuous nonlinear function and $B~$ is an $n\times n$ real constant
matrix.

For example, the Lorenz system
\begin{equation*}\label{lorenz}%
\begin{array}
[c]{cl}%
\overset{\cdot}{x}_{1}= & \sigma(x_{2}-x_{1}),\\
\overset{\cdot}{x}_{2}= & x_{1}(\rho-x_{3})-x_{2},\\
\overset{\cdot}{x}_{3}= & x_{1}x_{2}-\beta x_{3},
\end{array}
\end{equation*}
\noindent with $n=3$, $x_0\in \mathbb{R}^3$, $\sigma=10$ and $\beta=8/3$, can be expressed in the form \eqref{unu}-\eqref{doi}, if one considers $p:=\rho$ as the bifurcation parameter, where
\[
g(x)=\left(
\begin{array}
[c]{c}%
\sigma(x_{2}-x_{1})\\
-x_{1}x_{3}-x_{2}\\
x_{1}x_{2}-\beta x_{3}%
\end{array}
\right)  ,~~B=\left(
\begin{array}
[c]{ccc}%
0 & 0 & 0\\
1 & 0 & 0\\
0 & 0 & 0
\end{array}
\right).
\]

In \cite{mo3}, it is proved that by using the periodic time-varying PS algorithm, any attractor of a system modeled by \eqref{unu}-\eqref{doi} can be numerically approximated and, also, expressed as a convex combination of a set of attractors of the considered system \cite{eu5,eu6,mo3,asta6,asta0,asta7,asta00} (in \cite{asta00}, a circuit implementation of the PS algorithm is presented).

The algorithm is based on the averaging property of the system modeled by \eqref{unu}-\eqref{doi}, to which the PS is applied (the analytic proof of this property is given in two different ways in \cite{eu5} and \cite{eu6}).

As is known, the global attractor (‘global minimal B-attractor’, ‘global uniform attractor’ or
‘maximal attractor’), represents a major research topic in dynamical systems, especially within the context of partial differential equations (see, e.g. \cite{disi4}). From the definition, a global attractor is a compact and invariant set containing all the dynamics evolving from all possible initial conditions. In other words, it contains all solutions, including stationary solutions, periodic
solutions, as well as chaotic attractors, relevant to the asymptotic behaviors of the system. The term local attractor is sometimes assigned to attractors that are not global attractors.

\emph{In this paper, by attractor one understands the global attractor numerically approximated by a (unique) solution through a chosen initial condition \cite{eu3,disi2,Milnor,asta}}.

For a given parameter $p$ which is a function of initial condition $x_0$, the attractor may contain several local attractors, which only attract trajectories from a subset of initial conditions, specified by its basin of attraction (see e.g. \cite{eu3}).

Neural networks, the successors of analog computers, are useful in exploring the relationship between electrical activities and information transfer in the human brain to reveal the mechanisms of
memory, association, and emotional expression \cite{asta2,asta3}. Neural networks are systems that are typically used to recognize, describe, or even to control the dynamics.

A chaotic system modeled by \eqref{unu}-\eqref{doi} is the classical Hopfield
Neural Network (HNN) model, a highly significant model since it can emulate complex, dynamic behaviors, similar to how
the brain processes information \cite{asta4}
In the 1980s, Hopfield proposed the HNN in his pioneering works \cite{pio}\cite{pio2}, which combines statistical mechanics
and can be derived from an electronic circuit. Fractional-order HNNs are considered in \cite{asta5} and \cite{asta6}.

The HNN considered in this paper is a simplified popular variant of the simplest continuous-time model of a recurrent neural network that ``realizes everything'' \cite{every2} in the sense that its dynamics can be determined with an arbitrary accuracy using its parameters, and presents a``maximal dynamic complexity'' \cite{every}, which is modeled by the following Kirchhoff equations \cite{pio2,yang} (see also \cite{asta0,pul1,pul2}):

\begin{equation*}\label{eq}
\dot{x}_i=-p_ix_i+\sum_{j=1}^3w_{ij}f(x_j),\quad i=1,2,3,
\end{equation*}

\noindent where $f$ is a monotone function, $f(x_j)=\tanh(x_j)$ is a continuous approximation of the inherent discontinuity related to HNN, $p_i\in \mathbb{R}^+$ are proportional constants\footnote{Commonly, $p_i$ in HNN models are set to one.} and the weights $w$ considered in this paper are defined by the weight matrix $W$
\begin{equation}\label{coe}
W=
\begin{pmatrix}
    w_{11}& w_{12}& w_{13}\\
    w_{21}& w_{22}& w_{23}\\
    w_{31}& w_{32}& w_{33}
\end{pmatrix}
=
\begin{pmatrix}
2&-1.2&0\\
1.9995& 1.71&1.15\\
-4.75&0&1.1
\end{pmatrix}.
\end{equation}
Here, set $p_2=p_3=0$ and $p_1:=p$ as the bifurcation parameter.
With these ingredients, the considered HNN model has the following form:
\begin{equation}\label{primu}
\begin{array}
[c]{cl}%
\overset{\cdot}{x}_{1}= & -px_1+2\tanh x_1-1.2\tanh x_2,\\
\overset{\cdot}{x}_{2}= & -x_2+1.9995\tanh x_1+1.71\tanh x_2+1.15\tanh x_3,\\
\overset{\cdot}{x}_{3}= & -x_3-4.75\tanh x_1+1.1\tanh x_3,
\end{array}
\end{equation}
whereas, the elements \eqref{unu}-\eqref{doi} are

\begin{equation*}
\begin{array}
[l]{l}%
B=\begin{pmatrix}
1&0&0\\
0& 0&0\\
0&0&0
\end{pmatrix}\\
\end{array}
g(x)=
W\begin{pmatrix}
\tanh x_1\\
\tanh x_2\\
\tanh x_3
\end{pmatrix}=
\begin{pmatrix}
2\tanh x_1-1.2\tanh x_2\\
-x_2+1.9995\tanh x_1+1.71\tanh x_2+1.15\tanh x_3\\
-x_3-4.75\tanh x_1+1.1\tanh x_3
\end{pmatrix}.
\end{equation*}

The hyperbolic tangent function $\tanh$ used in the HNN system can be
used to design multiscroll memristive neural networks with application to image encryption \cite{asta2}.

There are applications where transient chaos can be quite disastrous, such as the situation of species extinction or power-grid
voltage collapse.
As known, the notion of attractors as limit sets, as time goes to infinity. Numerically, this characteristic is specified as a sufficiently large time integration. Therefore, if after a relatively long but reasonable integration time, chaos vanishes, whereas the underlying transient should be called ``set'', not ``attractor'', which is an asymptotical approach. As shown in this paper and in \cite{asta0}, depending on the coupling coefficients $W$, the HNN system might have non-chaotic attractors but finite time chaotic transients.

Sometimes, the conversion of transient chaos into sustained chaos can avoid catastrophes
related to sudden chaos collapses (chaos
anticontrol) \cite{eu2}. However, there are situations where chaotic transients are
undesirable, for which chaos control techniques are useful.
A recently found phenomenon of transient chaos, doubly transient chaos, term used to emphasize the fact that in many systems, chaotic solutions finally stop
(chemical reactions, binary star behavior, etc.) and is likely far less predictable than previously thought and is fundamentally different from hyperbolic and nonhyperbolic transient chaos
\cite{eu1}.

Based on the connection of basins of attraction of attractors with their equilibria in the phase space, it is natural to suggest the following computationally attractors classification:
An attractor is called a self-excited attractor if its basin of attraction intersects with any open neighborhood of an equilibrium; Otherwise, it is called a hidden attractor. A particular case is represented by hidden chaotic attractors in systems without equilibria \cite{hid,hid2}.

In \cite{asta0}, is shown that the HNN system \eqref{primu} presents long hidden chaotic transient sets. In this paper, we concern with approximating the attractors of the HNN system and developed a new chaos control/anticontrol technique, based on the PS algorithm.

The manuscript is organized as follows: Section 1, Introduction; Section 2, where the PS algorithm is briefly described; Section 3, where it is shown how attractors can be decomposed as convex combination of other attractors of the considered system; Section 4, where the PS algorithm is applied to approximate attractors of the considered HNN system and also, to decompose HNN's attractors as convex combinations of other attractors and the last section, Conclusions and Discussion Section. In Appendix is presented the Matlab code, which allows to use randomly the PS in the case of the HNN system.

\section{Parameter Switching Algorithm}
In this section, the PS algorithm is briefly introduced (details can be found in the indicated references).

Throughout, the following reasonable, mild conditions for most chaotic
systems are imposed:

\vspace{3mm}
\noindent(\textbf{H}) Existence, uniqueness, and boundness of solutions are assumed, and there exist only hyperbolic equilibria.
\vspace{3mm}

Note that, instead of boundness, the dissipative property,
an essential prerequisite for the existence of global attractors, can be used
(see, e.g. \cite{disi1,disi2}). Therefore, if $f \in C^1$, the dissipativity
is ensured via the divergence (Gauss) theorem (see, Chapter 16.9 \cite{disi3}).

Let $\mathcal{A}$ be the set
of all attractors depending on parameter $p$ and $\mathcal{P\subset}\mathbb{R}$ be the set of the corresponding
admissible values of $p$. Because of the solution uniqueness, for a chosen $x_0$ for each $p\in \mathcal{P}$, there corresponds a unique attractor $A_p\in\mathcal{A}$. Let $\mathcal{P}_N=\{p_1,p_2,\ldots,p_N\}\subset\mathcal{P}$, $p_1<p_2<\ldots,p_N$, a finite ordered set of $N\geq2$ parameters, to which there corresponds the attractors set $\mathcal{A}_N=\{A_{p_1},A_{p_2},\ldots,A_{p_N}\}\subset {\mathcal{A}}$.

Because of the assumption \textbf{H}, $\mathcal{A}$ is a non-empty set (see, e.g.
\cite{Letellier}). Therefore, for the
considered class of systems, a bijection between the sets
$\mathcal{P}$ and $\mathcal{A}$ can be naturally used.
In this way, for any $p\in\mathcal{P}$, a unique attractor is defined and reversely.

The PS algorithm is a time-varying, periodically
switching the parameter $p$ within a chosen set $\mathcal{P}_N$, according to some rules, while the IVP is integrated with a convergent scheme for ordinary differential equations with a fixed step-size $h$, on nodes $nh$, $n=1,2,\ldots$.
Symbolically, the algorithm is expressed as the following scheme

\begin{equation}\label{regula}
  S:=[m_1\cdot p_1,m_2\cdot p_2,\ldots,m_N\cdot p_N],
\end{equation}
where $\mathcal{M}_N=\{m_1,m_2,\ldots,m_N\}$, $m_i\in \mathbb{Z}^+$, are the set of weights of $p_i$, $i=1,2,\ldots,N$, and `$\cdot$' indicates the number of integration steps for which $p$ receives the values $p_i$.

The scheme $S$ should be read as follows: while the attractor is numerically generated along the trajectory, the first $m_1$ integration steps of size $h$, i.e. for the $1st,2nd,\ldots,m_1th$ steps, the parameter $p$ is set $p_1$. Then, for the next $m_2$ steps, $(m_1+1)th,(m_1+2)th,\ldots (m_1+m_2)th$, $p$ receives the value $p_2$ and the integration continues until $p=p_N$ for the steps $(m_{N-1}+1)th,(m_{N-1}+2)th,\ldots,(m_{N-1}+m_N)th$. Next, the algorithm repeats on the next set of $N$ subintervals, and so on.
Thus, the time periodicity of $p$ switching is $T=\sum_{i=1}^{N} m_ih$.

Let $p^\circ$ be the average of the switching values of $p$:

\begin{equation}\label{p0}
p^\circ=\frac{\sum_{i=1}^{N} m_ip_i}{\sum_{i=1}^N m_i}.
\end{equation}

Denote by $y_n$ the \emph{switched solution }generated by the PS algorithm with scheme \eqref{regula} and by $x_n$ the \emph{average solution }obtained by integrating the IVP with $p$ replaced with $p^\circ$ given by \eqref{p0}.

\begin{theorem}\cite{eu5,eu6}\label{th1}
The solution $y_n$ is a convergent approximation of the solution $x_n$.
\begin{proof} (Sketch of the proof).

By using a convergent numerical scheme with fixed step size $h$ to integrate the IVP \cite{eu6}, it can be proved that the global approximation error, $e_n$, between the switched $y_n$ and averaged solution $x_n$, is
\[
e_n\leq nh\lVert B \rVert \lVert x_0 \rVert \sum_{i=1}^N m_i|p_i-p^\circ|+\mathcal{O}(h).
\]

\end{proof}
\end{theorem}

Another proof of the convergence of the PS algorithm is based on the averaging theory  \cite{eu5}. Using the order function \cite{mo2}, it can be shown that the solution of the switching equation
\begin{equation*}\label{mao1}
\dot{{x}}(t)=p(t/\lambda)Bx(t)+g(x(t)),~~x(0)=x_0,
\end{equation*}
where $\lambda$ determines the switching moments and $p$ is a piecewise periodic function of period $T$ with averaged value of parameters switches denoted by $p'$
\[
p'=\frac{1}{T}\int_t^{t+T}p(u)du,
\]
is approximated by the solution of the averaged equation
\begin{equation*}\label{mao2}
\dot y(t)=p'Bx(t)+g(x(t)),~~ y(0)=y_0.
\end{equation*}

\begin{remark}\label{remi}
\hfill
\begin{enumerate}[itemsep=-1mm]
\item[i)]The initial condition $y_0$ for $y_n$ in Theorem \ref{th1} could be different from $x_0$, but within the same attraction basin;
\item[ii)]The selection of the sets $\mathcal{P}_N$ and weights $\mathcal{M}_N$ to obtaining a desired value $p^\circ$ is not unique;
\end{enumerate}
\end{remark}

Call the attractor generated by the average solution as the \emph{average attractor} denoted $A_{p^\circ}$, and the attractor generated by the PS algorithm as the \emph{switched attractor} denoted $A^*$.
\begin{theorem}\label{th2}\cite{mo3,eu5,eu6}
Given the set $\mathcal{P}_N$, with weights $\mathcal{M}_N$, for $N\geq2$, the switched attractor $A^*$ obtained with the PS algorithm represents a convergent approximation of the average attractor $A_{p^\circ}$.
\end{theorem}

It can be proved that $A^*$ and $A_{p^\circ}$ are attractors belonging to the set of attractors of the considered system \cite{mo3}.

The measure of the “success” of the attractor approximation, proved analytically by Theorem \ref{th2}, can be verified numerically too.
However, this task is usually a difficult one, because the size and shape of an attractor might change once with the
control parameter. Also, attractors geometric structure in $\mathbb{R}^n$ can
be complicated.
Considering the difficulties to comparing attractors, the following criterion, which is more practical rather than rigorous theoretical, and which is a suitable modification and adaptation of the concept of topological equivalence
\cite{mo1} is
introduced (see also \cite{Christy}, \cite{Fiedler}, \cite{mo1},
\cite{Kapitanski}, \cite{Letellier}):

\begin{criterion}
An attractor $A_{p_i}$ is considered to approximate numerically the attractor $A_{p_j}$, for $i\neq j$, if their geometrical forms in the phase space (almost) coincide and the sense of motion along the attractors is preserved.
\end{criterion}
\noindent Theorem \ref{th2} implies the following results \cite{mo3}.

\begin{corollary}\label{coro1}
Given the set $\mathcal{P}_N$ and weights $\mathcal{M}_N$, $N\geq2$, by using the PS algorithm one obtains the switched attractor $A^*$, which approximates the averaged attractor $A_{p^\circ}$.
\end{corollary}
\begin{corollary}\label{coro2}
Given an attractor $A_{p^\diamond}$ there exists a set $\mathcal{P}_N$ and a set of weights $\mathcal{M}_N$, $N\geq2$, such that $p^\diamond$ can be obtained from \eqref{p0}. Moreover, $A_{p^\diamond}$ is approximated by the switched attractor $A^*$ obtained with the PS algorithm.
\end{corollary}

To verify numerically the approximation given by Theorem \ref{th2},
Hausdorff distance, time series, hystograms, phase plots, have been computed for several examples (see e.g. \cite{asta7}).

Even though one of the convergence proofs of the PS algorithm is not related to some numerical scheme \cite{eu5}, the explicit numerical approach of the proof in \cite{eu6} allows the implementation of the algorithm via some fixed step-size scheme (in this paper, the Runge-Kutta method (RK4)) with the integration step $h$.

\begin{figure}[ptbh]
\centering
\includegraphics[width=0.6\columnwidth]{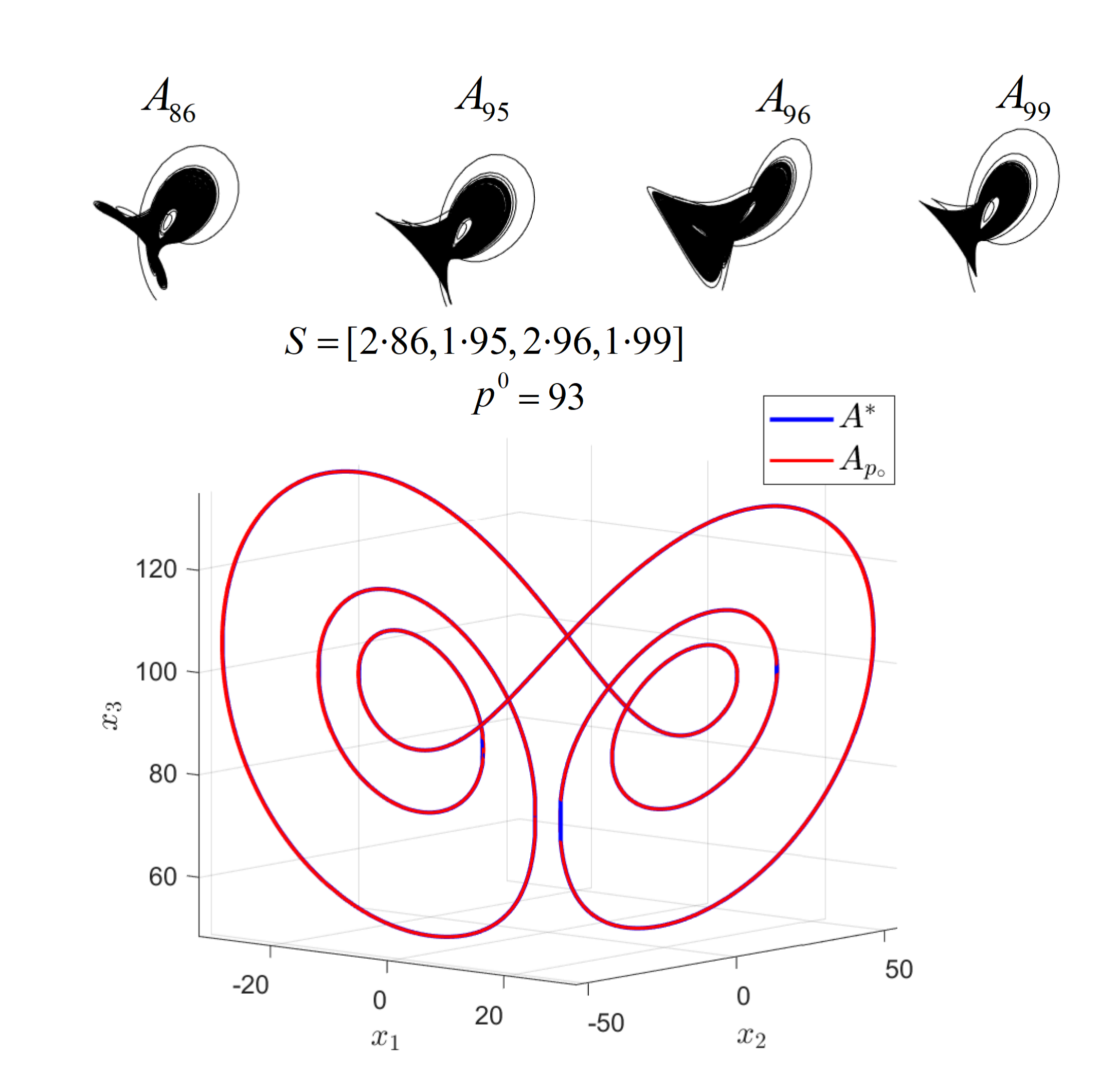}
\caption{Approximation of the attractor $A_{93}$ of the Lorenz system in the phase space, by using the PS algorithm with the scheme \eqref{regula}, for $P_4=\{86,95,96,99\}$, $m_1=2,m_2=1,m_3=2,m_4=1$, } $S=[2\cdot 86,1\cdot95,2\cdot96,1\cdot99]$. The switched attractor, $A^*$ (blue plot) approximates the average attractor $A_{p^\circ}$ (red plot), which is $p^\circ=93$. The plot reveals the match between the two attractors (transients are discarded). Attractors corresponding to the used parameters $86,95,96$ and $99$ are plotted on the top.
\label{fig1}%
\end{figure}

\begin{figure}%[ptbh]
\centering
\includegraphics[width=0.6\columnwidth]{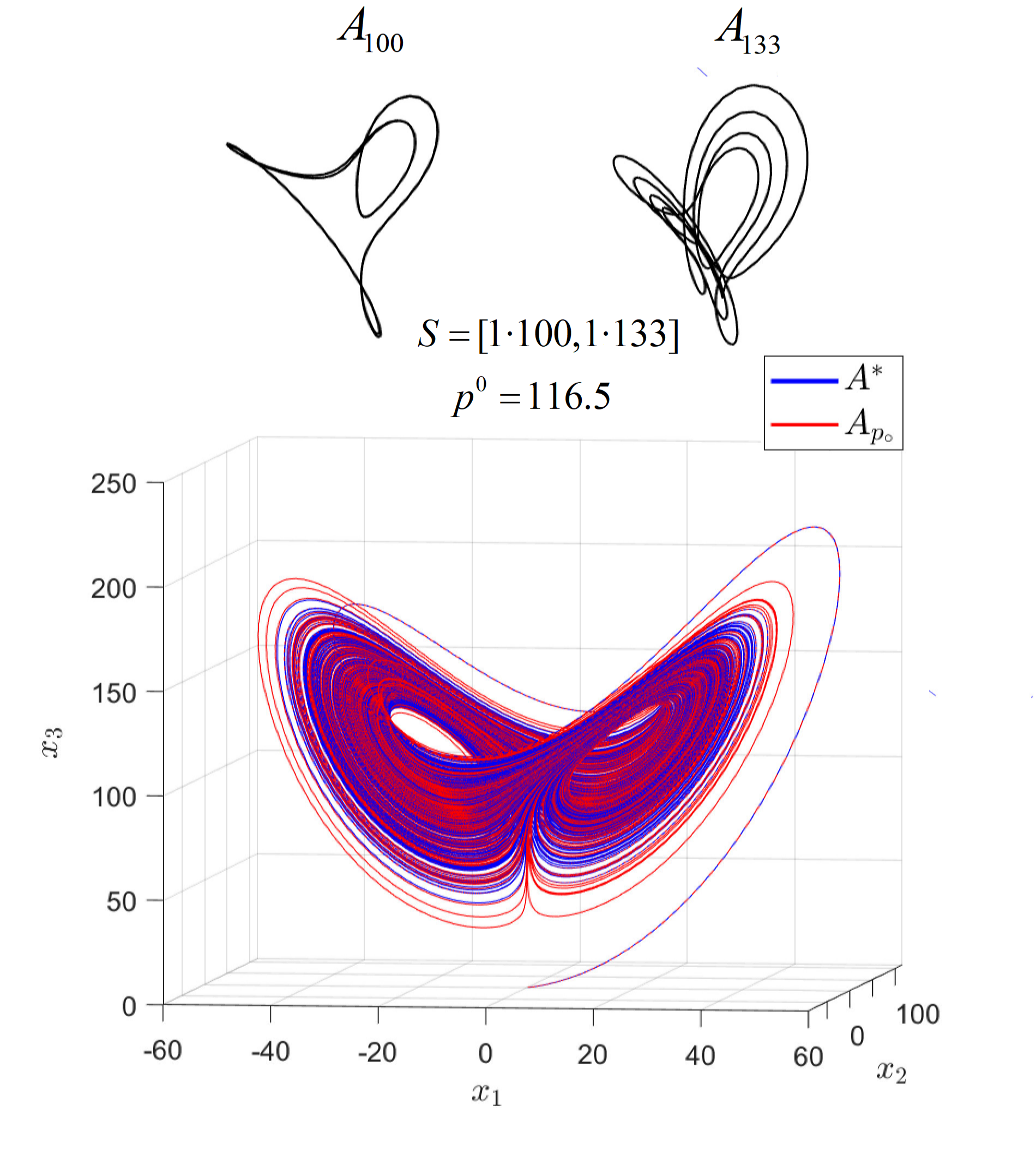}
\caption{Approximation of the chaotic attractor $A_{116.5}$ of the Lorenz system in the phase space, by using the PS algorithm with the scheme $S=[1\cdot100,1\cdot133]$ (i.e. parameter alternation). The average parameter value $p^\circ=116.5$. The used alternating parameters $p_1=100$ and $p_2=133$, generate stable cycles (top of figure).}
\label{fig2}%
\end{figure}

\begin{enumerate}
\item\label{ex1}
Consider the case of the Lorenz system and the set $\mathcal{P}_4=\{86,95,96,99\}$ with weights $\mathcal{M}_4=\{2,1,2,1\}$. Then, from Corollary \ref{coro1}, using scheme \eqref{regula}, $S=[m_1\cdot p_1,m_2 \cdot p_2,m_3\cdot p_3,m_4\cdot p_4]=[2\cdot86,1\cdot95,2\cdot96,1\cdot99]$, from \eqref{p0} one obtains $p^\circ=(m_1\times p_1+m_2\times p_2+m_3\times p_3+m_4\times p_4)/ (m_1+m_2+m_3+m_4)=93$, with the underlying averaged attractor $A_{p^\circ}:=A_{93}$ (stable cycle, blue plot in Fig. \ref{fig1}). The scheme $S$ reads as follows: For the first 2 integration steps, $p$ is set to $86$, the next step, $p=95$, and the next 2 steps, $p=96$, and so on, until the integration interval $I$ is covered. Applying the PS algorithm, one obtains the switched attractor $A^*$ (red plot in Fig. \ref {fig1}). Transients of both attractors have been removed. As can be seen, there is a perfect match between the two attractors.
\item\label{ex2}
For a given set $\mathcal{P}_N$, there exist several $N$ weights $m_i$, which give $p^\circ$ (Remark \ref{remi} ii)). Moreover, a specified value $p^\circ$ can be obtained with several sets of $\mathcal{P}_N$ and weights $\mathcal{M}_N$ with $N\geq2$. For example, from Corollary \ref{coro2}, the attractor $A_{93}$ ($A_{p^\diamond}$ in Corollary \ref{coro2}) can be approximated with the scheme $S=[1\cdot92,1\cdot94]$, or $S=[2\cdot86,1\cdot95,3\cdot97]$, which both give via \eqref{p0} $p^\diamond:=p^\circ=93$.
\item
Not only stable cycles can be approximated with the PS algorithm, but chaotic attractors too. For example, by switching (alternating) the parameter $p$ within the two elements set $\mathcal{P}_2=\{100,133\}$, with weights $m_1=m_2=1$, i.e., within every integration step, $p$ is alternated between the values of $100$ and $133$, one obtains the switched attractor $A^*$ (Fig. \ref{fig2}), which approximates the average chaotic attractor $A_{p^\circ}$, with $p^\circ=(100+133)/(1+1)=116.5$.
\end{enumerate}
The simple Matlab code presented in \cite{mo3} allows to generate via the standard RK method, the switched $y_n$ and averaged $x_n$ solutions, for a given system, function on the sets $\mathcal{P}_N$ and $\mathcal{M}_N$, initial conditions $x_0$ and $y_0$, and the integration step size $h$. In this paper, for the Lorenz system, the set of initial conditions $x_0$ and $y_0$ of the switched solution and average solution, are $(1,1,1)$ and $(1.01,1.01,1.01)$, respectively.

\section{Attractor decomposition}
In this section, it is shown how the PS algorithm can be used to decompose attractor function of other attractors.
%Analyzing the dynamics of the considered attractors to approximate the attractors $A_{93$ and $A_{116.5}$,
Let the sets $\mathcal{P}_N$ and $\mathcal{M}_N$, $N\geq2$.

\begin{proposition}\label{p7}
The average parameter value $p^\circ$ given by \eqref{p0}  is a convex combination of the elements of the set $\mathcal{P}_N$.
\begin{proof}
Denote
\begin{equation}\label{pnou}
\alpha_i=m_i/\sum_{k=1}^Nm_k, ~i=1,2,\ldots,N.
\end{equation}
 Then, the relation \eqref{p0} becomes
\begin{equation}\label{noua}
p^\circ=\sum_{k=1}^N\alpha_kp_k,
\end{equation}
with $ \sum_{k=1}^N\alpha_k=1,~  0<\alpha_k<1$.
\end{proof}
\end{proposition}
Next, a linear (bijective) order-preserved function between the sets $\mathcal{P}$ and $\mathcal{A}$,
\[
H:\mathcal{P}\rightarrow \mathcal{A},~~H(p)=A,
\]
 can be useful \cite{mo3}. Also, as proved in \cite{mo3}, these sets are order isomorphic. Therefore, over the field $\mathbb{R}^+$, on the set $\mathcal{A}$, two binary operations $(\mathcal{A},\oplus,\odot,\mathbb{R}^+)$, with $\oplus$ called hereafter \emph{addition of attractors }and $\odot$ \emph{multiplication of attractors by positive real numbers}, can be useful.

Then, attractors decomposition can be formulated.
\begin{corollary}\cite{mo3}\label{coro3}
Consider the sets $\mathcal{P}_N=\{p_1,p_2,\ldots,p_N\}$, weights $\mathcal{M}_N=\{m_1,m_2,\ldots,m_N\}$, $N\geq2$, and the attractors $A_k=H(p_k)$, $k=1,2,\ldots,N$.
The average attractor $A_{p^\circ}$ determined by $p^\circ$ can be expressed as
\begin{equation}\label{desc}
A_{p^\circ}=\alpha_1\odot A_1\oplus\alpha_2\odot A_2\oplus\ldots\alpha_N\odot A_N,
\end{equation}
where $\alpha_k=m_k/\sum_{k=1}^Nm_k$, $\sum_{k=1}^N\alpha_k=1$, $0<\alpha_k<1,~k=1,2,\ldots,N$.
\begin{proof} (Sketch of the proof) In \cite{mo3}, it is proved that one way to define $\oplus$ and $\odot$ is via the function $H$ as follows
\[
\alpha\odot A:=H(\alpha H^{-1}(A)),~~A\in \mathcal{A}_N,~\alpha\in \mathbb{R}^+
\]
and
\[
A_1\oplus A_2:=H(H^{-1}(A_1)+H^{-1}(A_2)).
\]
Next, apply $H$ to \eqref{noua}.
\end{proof}
\end{corollary}
Note that numerically, the relation \eqref{desc} means that the attractor $A_{p^\circ}$ is approximated by the switched attractor $A^*$, obtained by the PS algorithm.

In \cite{mo3} is proved that the operation $\oplus$ is commutative and associative. As a consequence, the order of the considered parameters in the scheme \eqref{regula} is not important.

Relation \eqref{desc} refers to the spanning of a set of vectors $v_i$, $i=1,2,\ldots,m$, in the space $\mathcal{V}$ over a field $\mathcal{F}$ (such as $\mathbb{R}$) where, for any vector $v\in \mathcal{V}$ there exist scalars $c_i$, $i=1,2,\ldots,m$, such that $v=\sum_{i=1}^m c_iv_i$.

Consider the Lorenz system in Example \ref{ex1}. The attractors generated by the set $\mathcal{P}_4$, $\mathcal{A}_4=\{A_{86},A_{95},A_{96},A_{99}\}$, are  plotted in Fig. \ref{fig1} (a). The switched attractor $A^*$ approximates the average attractor $A_{p^\circ}$ with $p^\circ=93$. With $A_{86}=H(86)$, $A_{95}=H(95), A_{96}=H(96), A_{99}=H(99)$, and $\alpha_k$, $k=1,2,3,4$, given by \eqref{pnou}, $\alpha_1=m_1/\sum_{k=1}^4 m_k=2/6=1/3,\alpha_2=1/6,\alpha_3=1/3$ and $\alpha_4=1/6$. From the relation \eqref{desc}, one obtains

\[
A_{93}=\frac{1}{3}\odot A_{86}\oplus\frac{1}{6} \odot A_{95}\oplus\frac{1}{3}\odot A_{96}\oplus\frac{1}{6}\odot A_{99}.
\]
From Example \ref{ex2}, the same attractor $A_{93}$ can be decomposed as follows:
\[
A_{93}=\frac{1}{3}\odot A_{86}\oplus\frac{1}{6}\odot A_{95}\oplus\frac{1}{2}\odot A_{97}.
\]

\begin{remark}\label{remusx}
 %\begin{itemize}
 \cite{mo3} Giving the ordered set $\mathcal{P}_N=\{p_{min},\ldots, p_{max}\}$, using scheme \eqref{p0},  $p^\circ$ remains within the parameter interval $p^\circ\in(p_{min}, p_{max})$. Therefore, similarly, via $H$ the order induced by $\mathcal{P}_N$ in the set $\mathcal{A}_N=\{A_{min},\ldots,A_{max}\}$, implies that the switched attractor $A^*$, which approximates the average attractor $A_{p^\circ}$, verifies $A^*\in (A_{min},A_{max})$. Also, from the convexity of $p^\circ$, $A^*$ and $A_{p^\circ}$ are different from all attractors in $\mathcal{A}_N$.
%\end{itemize}
\end{remark}

\section{PS algorithm for attractors of the HNN system}
Consider the HNN system in the form \eqref{primu}.

\begin{figure}%
\centering
\includegraphics[width=0.6\columnwidth]{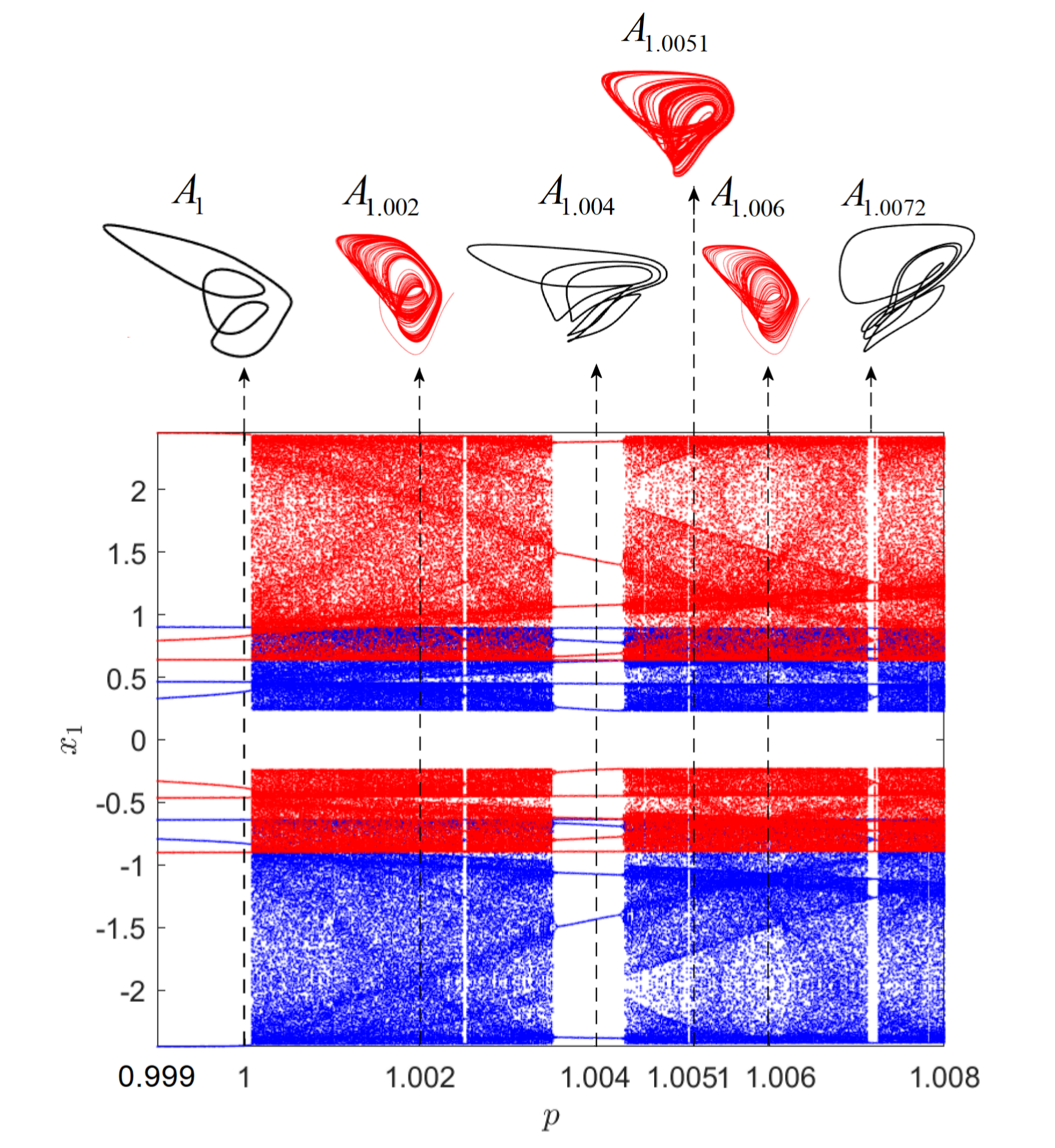}
\caption{Bifurcation of the HNN system for $p\in[0.999,1.008]$ of the component $x_1$, after transients are removed. Several chosen parameters with their corresponding attractors are indicated.}
\label{fig3}
\end{figure}

To facilitate the understanding of attractors approximation and decomposition of the HNN system, one can evaluate the bifurcation diagrams within a small neighborhood of $p=1$ (Fig. \ref{fig3}), generated with the same RK4 scheme as used for the PS algorithm.
The two colors, red and blue, indicate two symmetric with respect to the origin determined by the two equilibria. The initial conditions for the switched solution, and for the average solution, $x_0$ and $y_0$, respectively, are $(0.1,0.1,0.1)$ and $(0.2,0.2,.2)$.

\begin{figure}
\centering
\includegraphics[width=0.8\columnwidth]{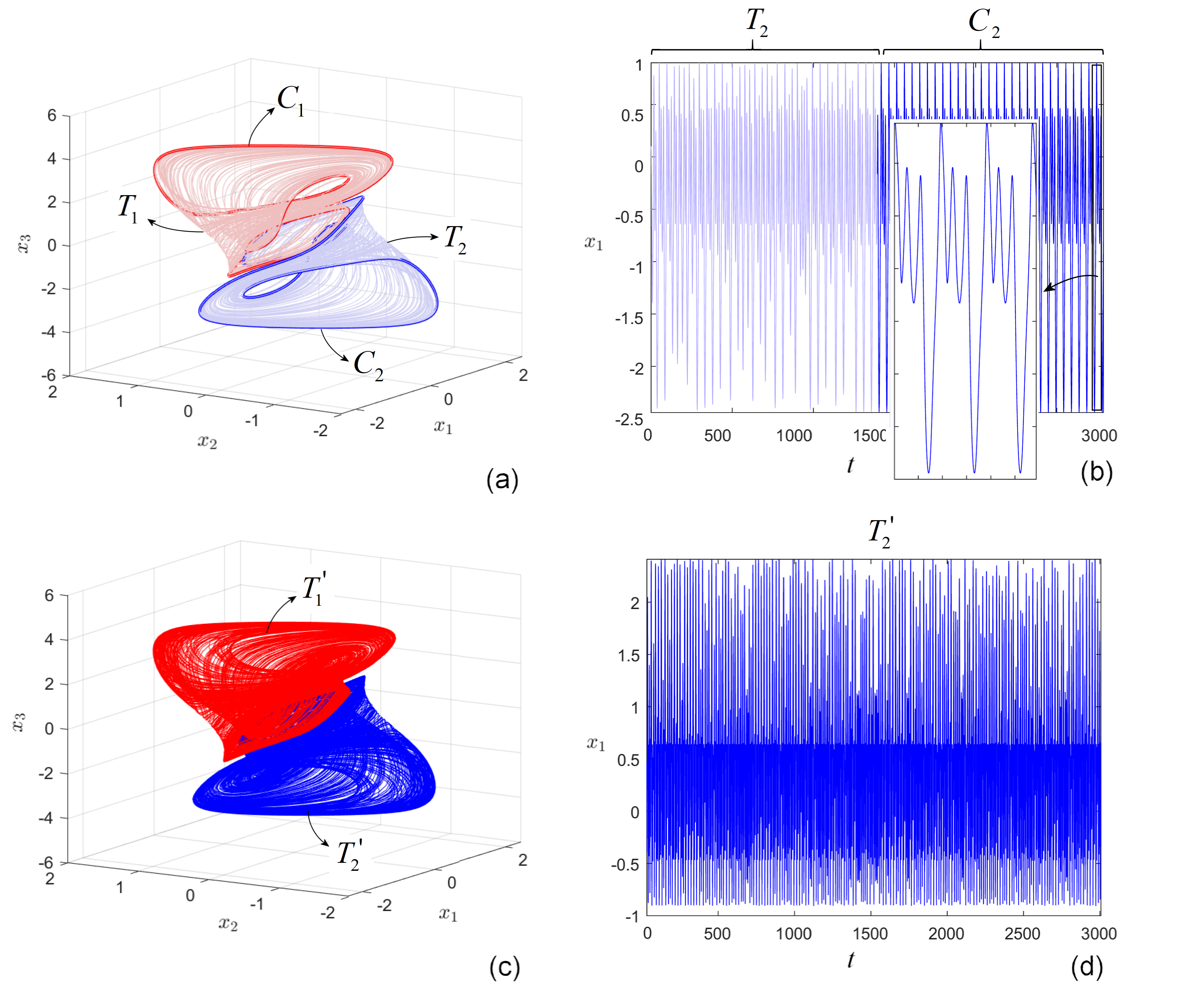}
\caption{(a) For $p=1$, the HNN system presents hidden chaotic transients $T_{1,2}$ (light red and light blue plot) which, after a relatively long time interval reach one of the two hidden attractors, stable cycles $C_{1,2}$ (thick red and blue plot); (b) Time series and zoomed detail of variable $x_1$; (c) Using the Matlab routine $ode45$ without options, the obtained result suggests a wrong answer for the of existence of chaotic attractors ($T_1'$ and $T_2'$); (d) Time series of the variable $x_1$.}
\label{fig4}%
\end{figure}

\begin{remark} (Hidden Chaotic Transients of HNN system \eqref{primu})\\
As shown in \cite{asta0} (where $p=1$), for the coefficients given by $W$ in \eqref{coe}, the HNN system has only chaotic transients, not attractors, which are proven to be hidden too \cite{asta0}. Thus, after a relatively long integration time the chaotic behavior of these two transients (denoted $T_{1,2}$ in Fig. \ref{fig4} (a), light red, and plot respectively) vanishes, and the trajectories reach one of the hidden stable cycles $C_{1,2}$ (red and blue thick plots in Fig. \ref{fig4} (a)). From the time series and zooming in Fig. \ref{fig4} (b), one deduces that the chaotic transient lasts till, about $t=1300$.
Note that if for $p=1$, one use the Matlab routine $ode45$ (utilized in several works on HNN systems), the obtained trajectories generate apparently chaotic attractors, which seems to verify numerically the asymptotic definition of attractors ($T_1'$ and $T_2'$ in Figs. \ref{fig4} (c), (d)). However, increasing the precision of the routine $ode45$ by using e.g. '$options = odeset('RelTol',1e-6,'AbsTol',1e-9)$', the trajectories reveal the behavior of the chaotic transients presented in Figs. \ref{fig4} (a), (b), as obtained by the RK4 method.
\end{remark}

The bifurcation diagram together, with intensive numerical simulations, suggest the following result (see \cite{asta0}, where the HNN is implemented for $p=1$)
\begin{conjecture}
For $p=1$, the HNN system with the coefficients \eqref{coe} admits only chaotic transients sets, no chaotic attractors.
\end{conjecture}

On the other side, extremely small perturbations of coefficients determine important changes in system dynamics such as the existence of chaotic attractors beside the chaotic transients \cite{asta0} . If the coefficients $W$ are slightly changed, the HNN system presents hidden attractors, which underlines the robustness of HNN systems.

\subsection{Approximation of HNN attractors}

Chaotic dynamics in neuronal systems might be useful, and even necessary. On the other side, there are situations when the emerging chaotic dynamics are undesirable \cite{skard}. In both situations, the PS algorithm could be used to give a satisfactory solution.

Consider the plausible situation when, for some objective reasons, the underlying parameter value of some desired attractor of the HNN system \eqref{primu} cannot be set thereby the numerical integration cannot lead to that attractor.
In these situations, via Theorem \ref{th2}, Corollary \ref{coro2} and Corollary \ref{coro3}, the PS algorithm provides the following result
\begin{theorem}\label{th3}
Every attractor of the HNN system \eqref{primu}, can be approximated by the PS algorithm.
\end{theorem}
To implement numerically Theorem \eqref{th3}, see Remark \ref{remusx}.

\begin{figure}%[ptbh]
\centering
\includegraphics[width=0.6\columnwidth]{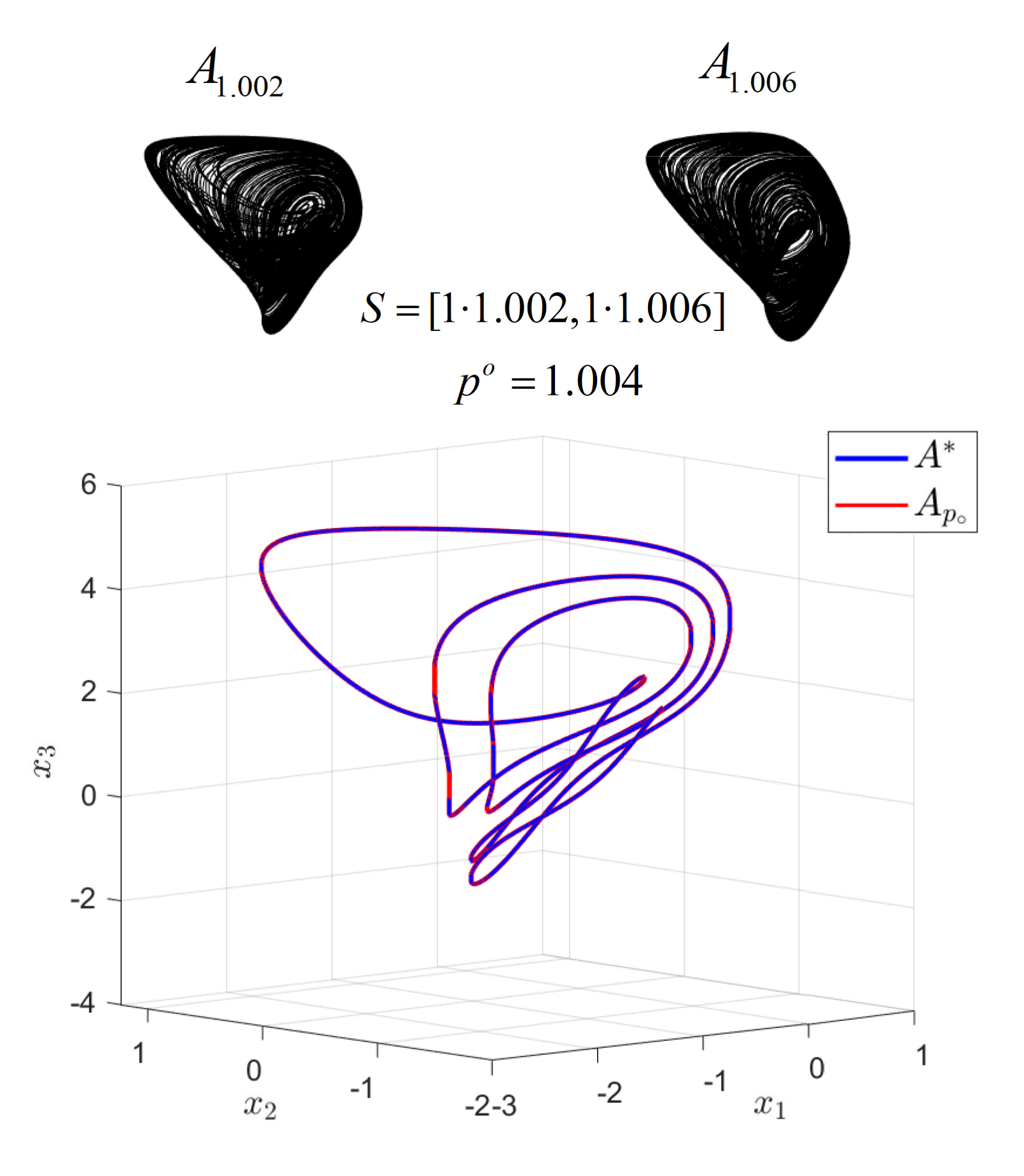}
\caption{Approximation of the stable cycle $A_{1.004}$ of the HNN system by using the PS algorithm with the scheme $S=[1\cdot1.002,1\cdot1.006]$. The averaged value $p^\circ$ is $p^\circ=1.004$ and the utilized attractors $A_{1.002}$ and $A_1.006$ are chaotic (plots in top of the figure).}
\label{fig5}%
\end{figure}

For example, to ``force'' the system to evolve, e.g. along the stable cycle $A_{1.004}$ (see Fig. \ref{fig5}) corresponding to $p=1.004$ (assuming this value cannot be set directly), one can consider the following scheme:

\begin{equation}\label{ee1}
S=[1\cdot1.002,1\cdot1.006],
\end{equation}
which means that the stable cycle $A_{p^\circ}$ with $p^\circ=1.004$ can be approximated by a switched attractor $A^*$ via alternating the parameter $p$ between the values $p=1.002$ and $p=1.006$ where in every integration the system evolves for $t\in I$.

Other more complicated PS schemes, which use more parameter values with other weights, can be used like the following one:
\begin{equation}\label{ee0}
S=[1\cdot1,2\cdot1.0016,2\cdot1.003,2\cdot1.0074,1\cdot1.008],
\end{equation}
which approximates the same attractor $A_{1.004}$.

The non-uniqueness of the scheme $S$ to approximate a desired attractor (e,g, schemes \eqref{ee1} and \eqref{ee0}), is important because it offers the possibility to choose among several variants to generate an approximation of the considered attractor.

Also, by looking to the way in which the PS algorithm acts and by considering the attractor decomposition \eqref{desc} determined by the algorithm, the similarity with chaos control or chaoticization becomes evident: by using the PS algorithm over a set $\mathcal{P}_N$, $N\geq2$, which generates chaotic dynamics, there exists a set $\mathcal{M}_N$ such that the obtained switched attractor approximates a regular attractor (chaos control-like action).

 On the other side, if one switches the parameter $p$ within a set $\mathcal{P}_N$, which generates regular dynamics, it is possible to find the weights $\mathcal{M}_N$ such that the obtained attractor is chaotic (anticontrol-like action).
 For example, if one uses the set of parameters that generate the stable cycles $A_{1}$, $A_{1.004}$ and $A_{1.0072}$ (see the bifurcation diagram in Fig. \ref{fig4}), with the scheme
 \begin{equation}\label{ee2}
 S=[1\cdot1,3\cdot1.004,4\cdot1.0072],
 \end{equation}
 one obtains the chaotic attractor $A^*$, an approximation of the attractor $A_{1.0051}$.

\subsection{HNN attractors as convex combinations of other attractors}\label{subsec}

As for the case of the Lorenz system, the decomposition \eqref{desc} can be used to express attractors of the HNN system function of other attractors.

 For example, starting from the scheme \eqref{ee1}, the attractor $A_{1.004}$ can be decomposed based on the attractors $A_{1.002}$ and $A_{1.006}$, as revealed by the following convex relation
\begin{equation}\label{ixus}
A_{1.004}=\frac{1}{2}\odot A_{1.002}\oplus\frac{1}{2}\odot A_{1.006},
\end{equation}
where $\alpha_1=\alpha_2=1/2$, $\alpha_1+\alpha_2=1$.

Suggested by scheme \eqref{ee0}, the same attractor $A_{1.004}$ can be expressed as the following convex combination:

\begin{equation}\label{desc_h}
A_{1.004}=\frac{1}{8}\odot A_{1}\oplus\frac{1}{4}\odot A_{1.0016}\oplus\frac{1}{4}\odot A_{1.003}\oplus\frac{1}{4}\odot A_{1.0074}\oplus\frac{1}{8}\odot A_{1.008},
\end{equation}
with  $\sum_{i=1}^5 \alpha_i=1$.

Note that, while the relation \eqref{ixus} expresses the decomposition of the stable cycle $A_{1.004}$ of two chaotic attractors, the relation \eqref{desc_h} shows that the stable cycle $A_{1.004}$ can be decomposed over the sets of one regular attractor (stable cycle $A_1$) and three chaotic attractors, $A_{1.0016}$, $A_{1.0074}$, and $A_{1.008}$.

Considering the chaotic attractor $A_{1.0051}$, decomposed by scheme \eqref{ee2}, one obtains the following relation
 \[
  A_{1.0051}=\frac{1}{8}\odot A_{1}\oplus\frac{3}{8}\odot A_{1.004}\oplus\frac{1}{2} \odot A_{1.0072},
  \]
which suggests that a chaotic attractor can be decomposed using regular attractors.

Moreover, a regular attractor, as $A_{1.004}$, can be expressed as combination of other regular attractors (such as the attractor $A_{1.004}$ obtained from the regular attractors $A_{1}$, $A_{1.072}$). Similarly, chaotic attractors can be decomposed as a combination of other chaotic attractors.

Theorem \ref{th2} can be generalized as follow
\begin{theorem}\label{tt2}
The switched attractor $A^*$ is a convergent approximation of the averaged attractor $A_{p^\circ}$, for randomly generated sets $\mathcal{P}_N$ and $\mathcal{M}_N$, $N\geq2$.
\begin{proof}
To prove Theorem \ref{tt2}, one considers the properties of the convex combination of $p^\circ$ (Proposition \ref{p7}).
\end{proof}
\end{theorem}

In Appendix, a short Matlab code  $PS.m$ is presented (a modified variant of the code presented in \cite{mo3}), runs the PS algorithm for a larger number $N$ of parameters, chosen randomly.

The code allows to verify that even for random values of $p$ and $m$ (chosen with respect to the algorithm requirements), the obtained results still sustain the analytical results. The code uses the function $rand$, which generates real numbers with a uniform distribution in the interval $(0,1)$.

\noindent The Matlab sequence to generate random $p$ within the range $[a,b]$ and $m$, $m<Mmax$ is

\vspace{3mm}
\begin{BVerbatim}
p=a+rand*(b-a),
m = ceil(Mmax * rand)
\end{BVerbatim}
\vspace{3mm}

The program asks the integration time $T_{max}$, the integration step size $h$, $N$, the interval $[a,b]$ within $p$ is randomly generated, the largest value of $m$, $Mmax$, $m<Mmax$, and the initial conditions $x_0$ and $y_0$. The outputs are the set $\mathcal{P}$, $\mathcal{M}$, $p_0$, and the averaged solution $x$ and the switched solution $y$.

\vspace{3mm}
\begin{BVerbatim}
[P,M,p0,y,x]= PS(Tmax,h,N,a,b,Mmax,x0,y0)
\end{BVerbatim}
\vspace{3mm}

\noindent For example, with the command

\begin{figure}[ptbh]
\centering
\includegraphics[width=0.7\columnwidth]{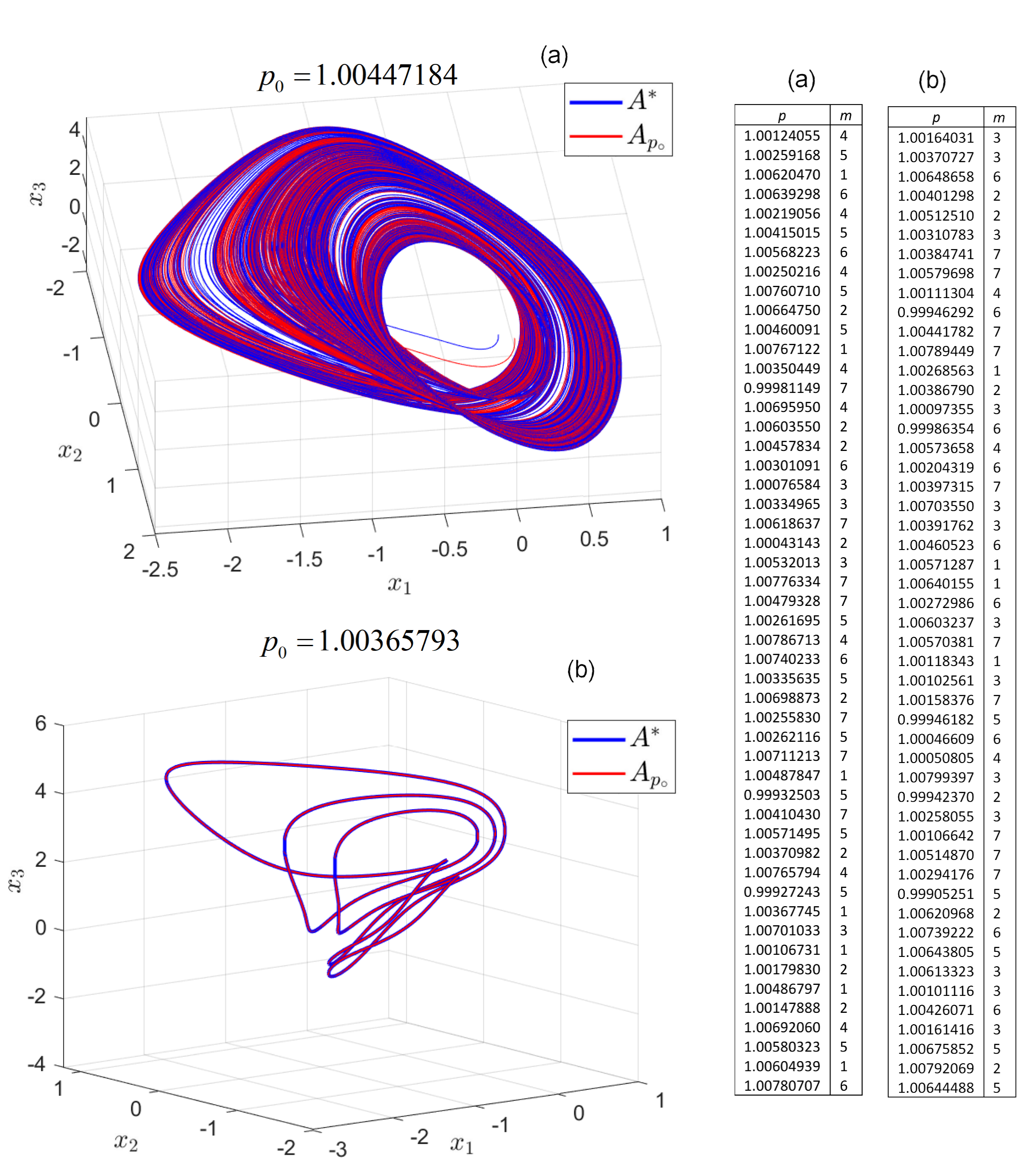}
\caption{The PS algorithm applied to approximate attractors of the HNN system with $50$ randomly chosen sets $\mathcal{P}_{50}$ and $\mathcal{M}_{50}$. (a) Overplot of the chaotic attractor $A_{1.0044718}$  and its   switched attractor; (b) Overplot of the stable limit cycle $A_{1.00365793}$ and its switched attractor. }
\label{rand}
\end{figure}

\vspace{3mm}
\begin{BVerbatim}
[P,M,p0,y,x]=PS(2500,0.005,20,0.999,1.008,7,[.1,.1,.1],[.2,.2,.2]),
\end{BVerbatim}
\vspace{3mm}

\noindent which generates 50 random parameter values within $[0.999,1.008]$ (corresponding to the bifurcation diagram in Fig. \ref{fig3}), with random weights $m<7$, the integration step size $h=0.005$, and initial conditions $[0.1,0.1,0.1]$ and $[0.2,0.2,0.2]$, one obtains the chaotic attractors in Fig. \ref{rand} (a), where the switched and average attractors are overplotted. The generated values of $p$ and $m$ are presented in Table (a) (with the first $8$ decimals displayed), which generates
\[
p_0=\frac{\sum_{k=1}^{50} m_kp_k}{\sum _{k=1}^{50}m_k}=1.00447184.
\]

\noindent Similarly, one can obtain the approximation of some stable cycle (see Fig. (b) and the values of $p$ and $m$ being drawn in Table (b)). In this case, $p_0=1.00365793$. Due to the lower probability to obtain values $p^\circ$ corresponding to periodic windows, compared to the larger chaotic windows in the parameter space (see the bifurcation diagram), a relatively larger number of runs of the code is required to obtain $p^\circ$ so that $A_{p^\circ}$ is a stable cycle.

\section{\label{sec4}Conclusions and Discussion}

In this paper, the PS algorithm is briefly introduced and is applied for the first time to approximate and decompose attractors of the HNN system \eqref{primu}. For the great majority of continuous chaotic systems, such as Lorenz, Roessler, Chen system etc., a chaotic system can be brought to the form required by the PS algorithm \eqref{unu}-\eqref{doi}, i.e., a linear dependence of a bifurcation parameter, $p$. It is underlined that, for the bifurcation parameter $p=1$, with the coefficients $W$ given by \eqref{coe}, the HNN system has only chaotic transient sets, not chaotic attractors. Also, by using the opportunity given by the PS algorithm, to decompose an attractor as a convex combination of a set of other given attractors, it is shown both analytically and numerically too, that every attractor of the HNN system can be considered as a convex combination of other attractors. Moreover, it is shown that this decomposition can be interpreted as a control-like or anticontrol-like technique: the PS algorithm allows to approximate a regular attractor or a chaotic attractor, respectively. The numerical scheme used to integrate the IVP is,  RK4. It is warned that, without $options$, the Matlab routine $ode45$ could give for this system wrong results concerning the chaotic attractors.
A future study will be carried out to analyze the potential algebraic structure defined by the two binary operations $\oplus$ and $\odot$ over the set of attractors of a dynamical system modeled by \eqref{unu}-\eqref{doi}.
\vspace{3mm}

\section*{Appendix}\label{apa}

\section*{Matlab code for randomly using  the PS algorithm}\label{cod}

The $PS.m$ Matlab code can be used to approximate randomly the set of $N\geq2$ parameters and weights used by the PS algorithm to approximate the attractors of the HNN system \eqref{primu}.  The code is a modification of the Matlab code presented in \cite{mo3}. A total of $N$ random parameter values within the range \texttt{[a,b]} and related weights within the range \texttt{[1,Mmax]} can be generated.  Here, as revealed by the bifurcation diagram \texttt{[a,b]=[0.999,1.008]}, and for \texttt{Mmax} it is recommended \texttt{Mmax<5-10}.

\vspace{3mm}
\emph{Input data:  }

\texttt{Tmax} integration interval;

\texttt{h}: step-size;

\texttt{N}: number of random generated parameters;

\texttt{a,b}: range for random \texttt{p};

\texttt{Mmax}: superior limit of random generated weights;

\texttt{x0, y0} initial conditions of the attractor to approximate $A_{p^\circ}$ and $A^*$;

\vspace{3mm}
\emph{Output data:}

\texttt{P}: the set of \texttt{N} random parameters \texttt{p};

\texttt{M}: the set of the random weights of \texttt{p};

\texttt{p0}: the average \texttt{p};

\texttt{y}: the switched solution;

\texttt{x}: the average solution.

\vspace{3mm}
\emph{Example}:

\begin{verbatim}
[P,M,p0,y,x]=PS(2500,0.005,50,0.999,1.008,7,[.1,.1,.1],[.2,.2,.2]);
\end{verbatim}

% the approximating attractor A^*, respectively (column vectors);
%
% Output:
% p0: averaged parameter vale;
% y: A^0 vector;
% x: A^* vector;
%
% Example:
% [P,M,p0,y,x]=PS(2500,0.005,20,0.999,1.008,7,[.1,.1,.1],[.2,.2,.2]);
%

%\lstset{language=Matlab,%
%    breaklines=true,%
%    morekeywords={matlab2tikz},
%    keywordstyle=\color{blue},%
%    morekeywords=[2]{1}, keywordstyle=[2]{\color{black}},
%    identifierstyle=\color{black},%
%    stringstyle=\color{mylilas},
%    commentstyle=\color{mygreen},%
%    showstringspaces=false,%without this there will be a symbol in the places where there is a space
%    numbers=left,%
%    numberstyle={\tiny \color{black}},% size of the numbers
%    numbersep=9pt, % this defines how far the numbers are from the text
%}

%\lstset{style=Matlab-editor}

%\begin{multicols}{2} %\raggedcolumns

\begin{lstlisting}[belowskip=0pt]%[    basicstyle=\footnotesize, %or \small or \footnotesize etc.]

function [P,M,p0,y,x]=PS(Tmax,h,N,a,b,Mmax,x0,y0)

n=round(Tmax/h);% number of integration steps;
x=zeros(3,n);%3 is the system dimension in the phase space for HNN system
y=zeros(3,n);
i=1;j=1;
P=[];M=[];
for k=1:N %Sequences for N random 'p' and 'm'
    p=a+rand*(b-a);
    m=ceil(Mmax*rand);
    M=[M m];
    P=[P p];
end
%Switched attractor A^*, 'y'
l=length(P);
y(:,1) = y0;
while j<n
    for k=1:M(i)
        K_1= h*f(y(:,j),P(i));	
        K_2= h*f(y(:,j)+1/2*K_1,P(i));	
        K_3= h*f(y(:,j)+1/2*K_2,P(i));	
        K_4= h*f(y(:,j)+K_3,P(i));	
        y(:,j+1)=y(:,j)+1/6*(K_1+2*K_2+2*K_3+K_4);
        j=j+1;
    end
    i=mod(i,l);
    i=i+1;
end;
%Average attractor A^0, 'x'
p1=cumprod([M;P]);
p0=sum(p1(2,:))/sum(M);%p^0
j=1;
x(:,1) = x0;
while j<n %RK4
    K_1=h*f(x(:,j),p0);	
    K_2=h*f(x(:,j)+1/2*K_1,p0);	
    K_3=h*f(x(:,j)+1/2*K_2,p0);	
    K_4=h*f(x(:,j)+K_3,p0);	
    x(:,j+1)=x(:,j)+1/6*(K_1+2*K_2+2*K_3+K_4);
    j=j+1;
end
function ff = f(u,p)% HNN system
ff=zeros(3,1);
ff(1)=-p*u(1)+2*tanh(u(1))-1.2*tanh(u(2));
ff(2)=-u(2)+1.9995*tanh(u(1))+1.71*tanh(u(2))+1.15*tanh(u(3));
ff(3)=-u(3)-4.75*tanh(u(1))+1.1*tanh(u(3));

\end{lstlisting}
%\end{multicols}

%\begin{lstlisting}

%\tiny

%\begin{multicols}{2}
%\lstinputlisting[
%frame=single,
%numbers=left,
%style=Matlab-editor
%]{PS.m}
%\end{multicols}
\vspace{3mm}

%\bibliography{biblio}{}
%\bibliographystyle{plain}

\bibliographystyle{unsrt}
\bibliography{biblio}{}
%\bibliography{bibliography}

\textbf{Declarations}

The Matlab code is available from the corresponding author.

The authors declare that they have no conflict of interest.

Authors contributed equally to this work

\newpage

\end{document}